\documentclass[12pt]{article}
\usepackage[utf8]{inputenc}
\usepackage{amsmath,amssymb,latexsym}
\usepackage[font=small,labelfont=bf,   justification=justified,   format=plain]{caption}
\usepackage{subcaption}
\usepackage{adjustbox}
\usepackage{enumerate}
\usepackage{multirow}
\usepackage{graphics}
\usepackage{graphicx}
\usepackage{epsfig}
\usepackage{tikz}
\usepackage[square,sort&compress,comma,numbers]{natbib}
\usepackage{float}
\usepackage{fancyhdr}
\usepackage[a4paper, total={6in, 8in}]{geometry}
\usepackage{booktabs}
\usepackage{makeidx}
\usepackage{amssymb}
\usepackage{eurosym}
\usepackage{placeins}
\usepackage{url}
\usepackage{amsthm}
\usepackage{times}
\usepackage[T1]{fontenc}
\usepackage[utf8]{inputenc}
\usepackage{grffile}
\usepackage[super]{nth}

\graphicspath{{images/}}
\floatstyle{plaintop}
\restylefloat{table}

\newlength\imagewidth
\newlength\imagescale

\emergencystretch=\maxdimen
\begin{document}

\title{\bf Enhanced self-collimation effect by low rotational symmetry in hexagonal lattice photonic crystals}
\author{Zekeriya Mehmet Yuksel$^{1}$, Hasan Oguz$^{1}$, Ozgur Onder Karakilinc$^{2}$,\\ Mirbek Turduev$^3$\footnote{mirbek.turduev@manas.edu.kg}, Halil Berberoglu$^{4}$, Muzaffer Adak$^{1}$, Sevgi Ozdemir Kart$^{1}$\footnote{ozsev@pau.edu.tr}\\
   {\small $^1$ Department of Physics, Faculty of Science,  }\\
  {\small  Pamukkale University, 20160 Pamukkale, Denizli, Turkey}\\
  {\small $^2$Department of Electric Electronic Engineering, Faculty of Engineering,} \\
  {\small  Pamukkale University, 20160 Pamukkale, Denizli, Turkey} \\
  {\small $^3$ Electrical and Electronics Engineering, Faculty of Engineering, } \\
   {\small Kyrgyz-Turkish Manas University, Bishkek, Chuy, 720038 Kyrgyz Republic }\\
   {\small $^4$ Department of Physics, Polatli Faculty of Science and Letters, } \\
   {\small Ankara Haci Bayram Veli University, 06900 Polatli, Ankara, Turkey }\\
 }
 
\vskip 1cm
\date{}
\maketitle
 \thispagestyle{empty}

\begin{abstract}
  \noindent
  
  In this study, we present the design of a photonic crystal (PC) structure with a hexagonal lattice, where adjustments to the PC unit cell symmetry reveal an all-angle self-collimation (SC) effect. By optimizing opto-geometric parameters, such as the rotational angle of auxiliary rods and adjacent distances, we analyze the SC property in detail, leveraging group velocity dispersion (GVD) and third-order dispersion (TOD) characteristics. We also investigate the relationship between symmetry properties and their influence on dispersion characteristics. Through symmetry manipulation, we gain a comprehensive understanding of the underlying mechanisms governing light collimation and confinement in the proposed configurations. The PC structure with a $C_1$ symmetry group exhibits all-angle SC effect within the range of $a/\lambda=0.652$ and $a/\lambda=0.668$ normalized frequencies, with a bandwidth of $\Delta\omega/\omega_c = 2.4\%$. Further breaking the symmetry, transforming from $C_1$ to $C_2$ group symmetry, enhances the SC bandwidth to $\Delta\omega/\omega_c = 6.5\%$ and reveals the perfect linear equi-frequency contours (EFC) at two different frequency bands: all-angle SC between $a/\lambda=0.616$ and $a/\lambda=0.656$ normalized frequencies in the \nth{4} transverse magnetic (TM) band and between $a/\lambda=0.712$ and $a/\lambda=0.760$ in the \nth{5} TM band. Here, GVD and TOD values vary between $7.3 (a/2\pi c^2) - 254.3 (a/2\pi c^2)$ and $449.2 (a^2/ 4\pi^2c^3) - 1.3\times 10^5 (a^2/ 4\pi^2c^3)$ for the TM \nth{4} band, respectively. Also, GVD and TOD values vary between $(182.5 - 71.3) (a/2\pi c^2)$ and $((-24380) - (-9619)) (a^2/ 4\pi^2c^3)$ for the TM \nth{5} band values. Additionally, we propose a composite/hybrid PC structure resembling $C_2$ group symmetry, where two auxiliary rods are replaced by rectangular photonic wires with the same refractive index and width equal to the diameter of auxiliary rods. This hybrid structure exhibits an all-angle SC effect with an operating bandwidth of $\Delta\omega/\omega_c = 11.7\%$, displays near-zero GVD and TOD performance, and offers enhanced robustness against potential fabrication precision issues.

  \vskip 1.0cm
  \bigskip

  
\end{abstract}

\section{Introduction}

Photonic Crystals (PCs) are periodic optical structures that manipulates the light wave propagation according to the formed energy bands of the photons. The PC concept entered the literature in 1987 with the independent studies of E. Yablonovitch \cite{Yablonovitch:93} and S. John \cite{John:87}. Their pioneering study on photonic bandgaps (PBGs) have received great attention because the propagation of electromagnetic waves at certain frequencies can be either prohibited or allowed, regardless of the polarization and propagation direction of electromagnetic waves. These properties lead to many interesting optical phenomena and applications by the photonic structures such as optical filters, PBG-based mirrors, demultiplexers, logic gates, etc., \cite{koshiba2001,Tekeste:06}. In recent years, important properties such as negative refraction \cite{Cubukcu2003}, super-prism \cite{gumus2018}, slow light, and self-collimation (SC) effects \cite{Wu2011} have gathered great interest of researchers.  Periodic modulation of the optical medium provokes the emergence of the SC phenomenon under appropriate opto-geometric circumstances. The diffraction-sensitive Gaussian beam in free space propagates through the periodic medium without broadening and retaining its spatial profile allowing light to be propagated by confining in PCs without the requirement of using any waveguides. The use of self-conducting regions of PCs allows the use of more transmission frequencies than those offered by defect-containing waveguide structures. This feature paves the way for the efficient transportation of light and offers important optical applications such as on-chip optical interconnects \cite{sato2015}, beam steering  \cite{Chuang2011}, bending \cite{Wu2011}, and light splitting \cite{Yasa:17}. Furthermore, this effect can be used to combine, route, and guide light beams by minimizing optical power loss in communication systems \cite{Noori2015}.

 Recently, an on-chip waveguide based on the SC phenomenon has been demonstrated in two-dimensional (2D) PCs \cite{sato2015}, and several interesting applications, such as channel-less waveguiding, bending of light, beam splitter, analog-to-digital converter, optical switch, etc. have been introduced \cite{Noori2015:2,Prather2007}. SC features are beneficial in waveguides for beam orientation \cite{Yifeng2014}, beam splitters \cite{Feng2012,Ren2012}, polarization beam splitters \cite{Noori2017}, logic gates \cite{Christina2012}, switches \cite{Zhang:07,Wang:12}, and demultiplexers \cite{Turduev:2013,Zheng:2014}. Moreover, PC research is an active field which has many recent developments like metastructures \cite{metamat}, nanoparticle structures \cite{nanopart}, near-zero-index materials \cite{nzimat}, graphene based photonic applications \cite{graphenmat}, plasmonic structures \cite{plasmonics} and multi-functional structures \cite{multifunction}.

When the PC studies in the literature so far are examined, it is seen that most of them deal with the optical properties of the unit cells with a high rotationally symmetry. High rotationally symmetric PCs represent a perfect symmetry pattern \cite{Joannopoulos:08:Book}. The symmetrical perfection of these structures reduces the freedom of geometric adjustment of the unit cells. Therefore, this situation also limits the control over their optical properties. In high rotationally symmetric periodic structures, the SC property is seen in a single direction within the PC which
cannot be managed \cite{zhou2022,witzens2002,noori2018}. Low rotationally symmetric PCs are obtained by introducing additional components to the unit cell or changing the shapes of existing components. 
By exploiting the low symmetry and adjusting the orientations of the unit cells, the direction of the SC can be controlled. Low-rotational symmetric PC structures cause manipulation of the shape of the EFC efficiently and lead to the emergence of oblique SC transmission \cite{Turduev:12}. The geometric and structural diversities enrich the dispersion properties of these periodicmaterials because of symmetry reduction applied to PCs \cite{Gumus2019}. Therefore, these PCs play an active role in the design of optical applications and phenomena such as oblique and broadband SC \cite{GIDEN2013,Gumus18_2,Gumus_2020}, super prism \cite{gumus2018}, wavelength\cite{Gumus2019,oguz2023symmetry}, polarization splitter \cite{Yasa:17,Yasa_2017}, anisotropic zero refractive index \cite{He2015}, and sensor \cite{erim19,Giden:22}.

It is important to note that SC in PCs presents a unique method for light propagation, where beams travel without divergence over considerable distances, a stark contrast to the approach used in defect-based photonic crystal waveguides. This is attributed to the PC’s tailored photonic band structure, ensuring that the group velocity–or the direction of energy propagation–remains nearly uniform across a spectrum of wavelengths, thus
allowing light beams to retain their shape and direction. Unlike SC that achieves guiding of light inherently through the crystal’s dispersion properties, defect-based guiding hinges on the strategic introduction of a defect or line defect into the PC, creating guided modes within the photonic band gaps for total light confinement. 
SC boasts several notable advantages, including a broader operational frequency range, devoid of the limitations posed by the photonic crystal’s specific bandgap properties. Moreover, the absence of a need for defect introduction simplifies the fabrication process, offering a simpler and potentially more cost-effective manufacturing approach. 
Additionally, the reliance on the crystal’s natural properties renders self-collimating PCs more tolerant to structural imperfections and disorders, enhancing their durability and performance consistency. Perhaps most significantly, SC facilitates seamless integration of multiple optical functionalities on a single chip, circumventing the complexities associated with coupling different waveguides or interfacing waveguides with other optical devices. This integrated comparison underscores the superiority of SC over defect-based waveguides in bandwidth, ease of fabrication, robustness, and integration capabilities, positioning it as a superior technique for advancing optical technologies in telecommunications, optical computing, sensors, and on-chip optical interconnects.

In this study, we propose the design of PC structure with a hexagonal lattice by adjusting the unit cell symmetry to reveal all-angle SC effect. The main objective is to explore and manipulate various optical properties including group velocity dispersion (GVD), third-order dispersion (TOD), and SC. Also, the relationship between the symmetry properties of PCs and their influence on dispersion characteristics is investigated. By analyzing the potential for tuning the optical properties through symmetry manipulation, we ensure a comprehensive understanding of the underlying mechanisms that govern light collimation and confinement in the proposed configurations. It may pave the way for the development of advanced photonic devices that exhibit desired performances in terms of transmission efficiency, dispersion control, and SC. Additionally, the effects of variation in lattice geometries and unit cell configurations on EFCs as well as photonic band structures are analyzed to achieve the desired optical behavior in the proposed structures.

\section{Problem Definition and Methodology}

The geometric diversity resulting from the symmetry reduction applied to PCs enriches their dispersion properties as well as spatial domain characteristics  \cite{Giden14}. In general PCs with square/rectangular lattices are more widely explored and analyzed by using the EFCs approach (dispersion engineering) compared to the hexagonal structures with low symmetry unit cells \cite{Wu2011}.  Here, higher refractive index contrast, lower group velocity dispersion, and larger photonic gaps of the hexagonal structure make them more suitable for revealing the desired optical properties such as SC, super-prism, optical filter, and waveguide \cite{gumus2018,Noori2015:2,Prather2007} as compared with those of the square/rectangular lattice PCs. Moreover, the hexagonal structure has more symmetric point reduction which enhances the SC effect, compared to the square structure.  For these reasons, the PC structure arranged in a hexagonal lattice is preferred to design low-symmetry PC configurations exhibiting inherent all-angle self-collimation characteristics, in this study. There are a few studies on the SC properties of hexagonal structures based on EFC engineering in theliterature. Photonic bandgaps, waveguides, SC properties, and negative refraction in hexagonal PC structures have been investigated \cite{Jovanovic:2011,Grimmer2017}. Unfortunately, manipulating rotational symmetry in hexagonal lattice PCs to enhance their dispersion behavior remains blurred. It is reported that interesting and applicable optical properties can be obtained by tuning the geometry of low symmetry introduced in the PC unit cell \cite{Turduev:2013, GIDEN2013, Gumus2019}. To the best of our knowledge, this work presents a detailed analysis of all-angle SC characteristics of the low symmetric PC structures ordered in hexagonal lattices by using the EFC engineering approach for the first time.

In this study, the Plane Wave Expansion (PWE) method \cite{mpb, pwe} is applied to calculate the band structure (dispersion relations) of Bloch modes and EFCs of the proposed low symmetry PC structures.
Correspondingly dispersion analyses are performed by using freely available MIT photonic bands tool (MPB) \cite{Joannopoulos:08:Book}. Although PWE can obtain an accurate solution for the dispersion properties (propagation modes and photonic band gap) of a PC, it still has some limitations. Transmission spectra, field distribution, and back reflections cannot be extracted as they only consider propagation modes.
In the PWE method, Maxwell's equations are solved by taking account of an eigenvalue problem. Here, the angular frequencies of the modes in PCs stand for the eigenvalues, while the amplitude coefficients of the fields indicate the eigenvectors. Our simulations have been performed by using dimensionless normalized frequency values ($a/\lambda$). Because the properties of PCs can be scalable, our results can be used in real-world applications. If the lattice constant is taken as 1µm, the minimum rod radii is $100 nm$ which is very convenient for fabrication \cite{Kirihara11}.
Additionally, both dielectric function and fields are introduced in the reciprocal $k$-space by using the Fourier series \cite{rumpf2006}.

\begin{figure}[H]
    \centering
        \includegraphics[width=\textwidth]{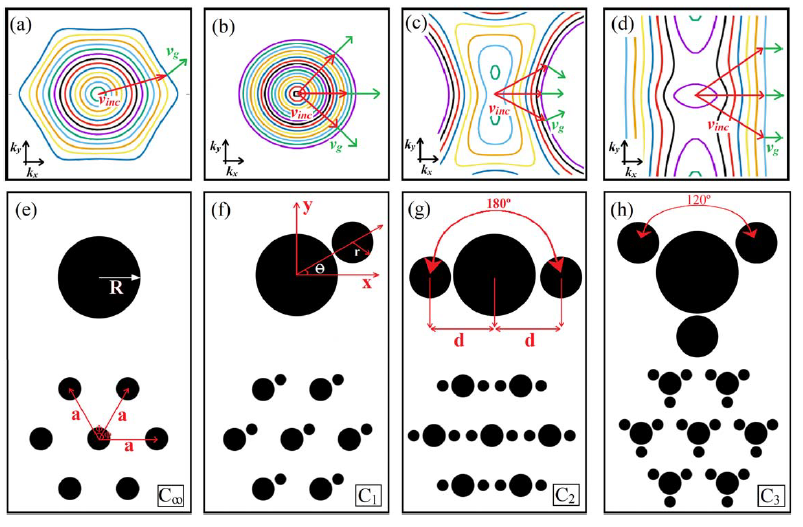}
    \caption{Wave propagation schemes under the analysis of EFC for (a) general arbitrary case, (b) homogeneous media, (c) medium exhibiting negative refraction effect, and (d) medium exhibiting self-collimation effect [29]. Schematic views of (e) hexagonal PC unit cell, (f) low rotational symmetric hexagonal PC unit cell with a single auxiliary dielectric rod ($C_1$ symmetric structure) having rotational angle of $\theta$, (g) two auxiliary dielectric rods ($C_2$ symmetric structure), and (h) three auxiliary dielectric rods ($C_3$ symmetric structure).}
  \label{fig:fig1}
\end{figure}

The projection of band structures $k$-space with constant frequencies can be expressed by EFCs.The curvature of EFCs in the $k$-space gives information/implications about the wave propagation behavior within/outside of the PC structure. Wave propagation schemes under analysis of EFC are represented in  Fig. \ref{fig:fig1}. 
As can be seen in Fig. \ref{fig:fig1}(a) the incident wave with velocity vector $\vec{v}_{inc}$(red arrow) leaves the PC interface perpendicularly to the group velocity $\vec{v}_{g}$(green arrow) due to the relation of $\vec{v}_{g}=\vec{\nabla}_k\omega(k)$ where $\omega(k)$is the angular frequency at the wave vector $k$.
Here, the group velocity of light represents the velocity of energy transport in the direction perpendicular to the EFCs within the PC structure. The schematic view of circular EFC is shown in Fig. \ref{fig:fig1}(b), where the light shows the wave propagation in an isotropic medium. In this case, the propagating wave does not undergo any diffraction. 
On the other hand, hyperbolic-shaped EFC produces negative refraction which results in a bi-refraction or focusing effect, as shown in figure \ref{fig:fig1}(c). Also, SC effect takes place when the wave passes through the flat EFC, as illustrated in Fig. \ref{fig:fig1}(d) \cite{DENG2021}. Here, the light of a specific wavelength propagates through a photonic structure with minimal diffraction and angular dispersion. EFC presents linear-shaped curves, as seen in  Fig. \ref{fig:fig1}(d) which indicates that the light propagates with a constant group velocity independent of incident light direction  \cite{kosaka1999}.
To evaluate and extract spatial field distributions of the propagating wave within the PC structure that are not within the scope of the PWE method, the Finite-Difference Time-Domain (FDTD) method is used in this study \cite{taf05,Lumerical}.
The PC structure ordered in hexagonal lattice is subjected to the periodic boundary conditions in two dimensions. Moreover, perfectly matched layers are applied to the simulation box to eliminate undesired boundary reflections \cite{Berenger1994}.
As known, the dispersion characteristics of photonic structures with isolated unit cells (dielectric rods in air) are more susceptible to transverse-magnetic (TM) polarization modes, while fully connected ones (air holes in the dielectric slab) provide rich dispersion properties for transverse-electric (TE) polarization modes \cite{Joannopoulos:08:Book}.
For this reason, throughout the study TM polarization mode is utilized to manipulate and tailor EFCs effectively.

general, the periodicity of a geometrical configuration is defined as the repetition of basic structural components/units (unit cells) in a specific order. Additionally, the periodic structure is examined in terms of the symmetry types in space. In this context, PCs can be represented by utilizing the rotational symmetry of the unit cells, which are periodically distributed in space. Rotational symmetry determines the degree at which the unit
cell must be rotated about its axis to reproduce the same structural view. The rotational symmetry group is symbolized by $C_{rot}$ where the subscript “$rot$” represents the ratio of $2\pi$ radians to the amount of angle that needs to be rotated for the object \cite{Grimmer2017}. The rotational symmetry groups analyzed in this study are depicted in Figs. 1(e) - 1(h). Here, Figs. 1(e) and 1(f) present high symmetric structure and $C_1$ symmetry group, respectively.  

It is also crucial to mention that the specific symmetry group of PCs significantly affects its EFCs, which are important in understanding the propagation of electromagnetic waves within the crystal. This relationship stems from the crystal’s geometric and dielectric properties, which determine the distribution of electric permittivity ($\epsilon$) across the unit cell and, by extension, influence the photonic band structures for different polarization modes. To analyze the effects of symmetry on EFCs, it’s important to consider how variations in the PC unit cell’s dielectric distribution impact the photonic band structures. This analysis can be done by using Maxwell’s equations within source-free dielectric environments, which govern the behavior of electromagnetic fields in the PCs. These fields can be expressed in terms of time-harmonic electric and magnetic field vectors, leading to a master equation that serves as an eigenvalue problem describing the photonic band structures. Furthermore, the electromagnetic field in periodic media can be expanded into a set of harmonic modes according to Bloch’s theorem \cite{mpb}. This expansion allows the representation of harmonic modes as a series of plane waves, which are crucial for understanding how light interacts with the PC structure. Here, the plane wave
expansion method simplifies Maxwell’s equations, enabling solutions for the photonic band structures and distinguishing between distinct polarization modes \cite{pwe}. Also, the Fourier coefficients for the dielectric permittivity matrix are calculated based on the area of the PC unit cell, highlighting how changes in the dielectric distribution ($\epsilon(r)$) directly affect the structure factor ($S(G)$) \cite{gumus2018,Giden14}. This factor depends on the geometrical parameters of the dielectric region within the unit cell, demonstrating the sensitivity of photonic band structures
to both geometric and dielectric properties of the lattice. Therefore, the specific symmetry group of a photonic crystal influences its EFCs by dictating the distribution of electric permittivity within the unit cell. This distribution, in turn, affects the structure factor and the resultant photonic band structures. By altering the symmetry group, one can manipulate the dielectric landscape of the PC unit cell, thus controlling light paths and behaviors within the crystal.

\section{Time and frequency domain analysis of self-collimation effect in hexagonal lattice with low symmetry PCs}

The values of the dielectric constants for the PC rods shown in Fig. 1(e) and the air background medium are taken as $\varepsilon_{PC} = 9.8$ and $\varepsilon_{air} = 1.0$, respectively. The rods’ radii are taken as $R = 0.2a$, where $a$ is the lattice constant of the structure.
The EFCs of \nth{1}, \nth{2}, \nth{3}, and \nth{4} bands for the TM polarization mode of the fully symmetric structure are created and presented in Figs. 2(a), 2(b), 2(c), and 2(d), respectively. As expected for the \nth{1} band, there are almost circular EFCs that exhibit the behavior of light propagation in a homogeneous medium.  Moreover, the other bands show more complex behavior at the EFC diagram. The SC effect is
particularly noticeable in the \nth{2} and \nth{4} bands at small impact angles. However, there isn’t any sign of the SC effect operating at wide incident angles. It reflects that one may obtain a special frequency response by tuning the lattice parameters (introducing a low symmetry effect into the PC unit cells) without breaking the structural lattice.
As known, the optical response of such structures is highly sensitive to the angular orientation
of the components within the PC unit cell \cite{Giden14}. Through dispersion engineering, this sensitivity can be used to obtain desired optical properties \cite{vuc01,Hamam2009,Zhou2008}.
For this reason, by introducing additional auxiliary dielectric rods having radii of $0.12a$ into fully symmetric hexagonal PC structure group $C_{\infty}$ is transformed into $C_1$ rotational symmetry group as represented in Fig. 1(f). The structural parameters of the PC structure with low symmetry are defined as the radii of the main dielectric rods ($R$), the radii of the auxiliary dielectric rods ($r$), and the distance between the centers of the main and auxiliary dielectric rods ($d$), as depicted in Figs. 1(e)-1(h).
Dispersion characteristics of all the configurations considered in this study are examined by using the PWE method. The inspections of the EFCs provide proper identification of the SC effect. Since all the opto-geometric parameters of PCs have strong effects on the appearance of linear-shaped EFCs, parametric optimizations are required to determine desired SC regions.
In addition, there should be enough spatial place for free adjustment of the auxiliary rods to analyze the impact of low symmetry within the PC unit cell and tailor of SC contours. All these pre-conditions force us to seek the optimum opto-geometric parameters. Optimizing opto-geometric parameters in PC structures involve fine-tuning various aspects of the crystal’s geometry to achieve desired optical properties. To find the optimum configurations, we have scanned all possible parameters within the unit cell of the structure by varying the diameter of the dielectric rods.
By considering these restrictions, the radius of hexagonal lattice PCs are fixed to the optimized value of  $R = 0.2a$ (this parameter is depicted in Fig. 1(e) as an inset). After taking the value of $R = 0.2a$, the other parameters are scanned to obtain SC effect. The radius of auxiliary rods $r$ and the distance $d$ are scanned from $0.09a$ to $0.18a$ and $0.31a$ to $0.39a$, with increments of $0.01a$ respectively. 

Finally, well-defined radii and distance parameters are as follows: $R = 0.2a$, $r = 0.12a$, and $d = 0.38a$, which leads to display all-angle SC behavior of $C_1$ symmetry group. It is important to note that we study the all-angle SC effect only for the case of rotational angle $\theta = 0^\circ$ of auxiliary PC rods, ensuring conventional SC with light propagating in a non-angled direction.

Figs. 2(e)-(h) show the EFC plots for the \nth{1}, \nth{2}, \nth{3}, and \nth{4} bands $C_1$ symmetric structure, respectively. The EFCs of the highly symmetric unit cell presented in Fig. 1(a) and Figs. 2 (a)-(d) exhibit an isotropic behavior. However, this isotropic behavior disappears because of the reduction in the unit cell symmetry, and different dispersion characteristics are revealed as seen in Fig. 1 (b)-(d), Figs 2 (e)-(h). This change, which is caused by the addition of an auxiliary dielectric rod to the unit cell structure, shows the significant effect reflected by symmetry reduction in PC. As expected, the \nth{1} band displays an isotropic medium effect, as shown in Fig. 2(a), but the other bands take different shapes according to the introduced auxiliary rods which reshape EFCs differently from those of a fully symmetric structure. The \nth{2} band has semi-linear contours with a wave-diverging effect (see Fig. 2(f)), while the \nth{3} band shows a tight-angle SC effect in the propagation direction (EFCs having a rectangular shape with curved corners) as shown in Fig. 2(g). On the other hand, the \nth{4} band exhibits the desired SC effect for all incident angles at the normalized frequency region of $a/\lambda=0.652$ and $a/\lambda=0.668$ as seen in Fig. 2(h). It can be reported that the analyzed structure can support the all-angle SC only at the narrowband with the bandwidth of $\Delta\omega/\omega=2.4\%$. Moreover, presented EFCs are not completely linear at these frequencies. To investigate the all-angle SC effect in time domain, the FDTD method is applied to the $C_1$ group symmetry PCs operating at the center frequency of $a/\lambda=0.664$. Steady-state electric-field intensity distributions generated by the incident waves at the angles of $0^\circ$ and $25^\circ$ are presented in Figs. 3(a) and 3(b), respectively. As seen from these figures, the engineered SC effect operates for a wide range of angles. To avoid an excessive number of shapes, only two angles are provided here. Since the four EFC lines ($a/\lambda=0.648$, $a/\lambda=0.656$, $a/\lambda=0.664$ and $a/\lambda=0.672$) located in the outermost layer of Figure 2(h) stay perpendicular to the $x$-axis throughout the entire structure, these frequencies can show the SC effect at all input angles. In addition, this situation is confirmed by creating steady-state electric field intensity distributions obtained by changing the angle of the source to the $x$-axis from $0^\circ$ to $80^\circ$ in $10^\circ$ increments. Here, regardless of the excitation angle of the incident wave, $C_1$ symmetry group PCs provide the propagation of light with negligible spatial dispersion.

It is also important to provide a qualitative analysis of the observed SC effect. Here, the quality of the SC effect is explored by utilizing GVD and TOD characteristics. GVD shows the dependence of group velocity on the optical wavelengths of light. GVD is defined as $\partial/\partial\omega(1/v_g)= \partial^2 k/\partial\omega^2$, where $k$ is the wave vector and $\omega$ is the angular frequency. 
It implies that the components of the incident signal propagate at different velocities at each wavelength, which leads to temporal pulse broadening. The positive sign of GVD indicates a normal distribution at which the group velocity decreases as the optical frequency increases, while the negative sign represents an anomalous distribution at which the group velocity rises with the optical frequency \cite{He2015}. TOD is described by the derivative of GVD with respect to the angular frequency given as $\partial^3 k/\partial\omega^3$. It affects the shape of the propagating pulse with a wide frequency spectrum. Regarding this matter, any PC structure that operates as a self-collimator should have as small as possible GVD and TOD values to avoid temporal dispersion and deformation of the propagated beam in a broad frequency range.
Figure 3(c) illustrates the behavior of calculated GVD and TOD values for the proposed structure as a function of the normalized frequency. 
It is observed that GVD and TOD values vary between $-200 (a/2\pi c^2)$ and $780 (a/2\pi c^2)$ and $104 (a^2/4\pi^2 c^3)$ and $18 \times 10^4 (a^2/4\pi^2 c^3)$ for the TM \nth{4} band, respectively. 
As we are concentrated on the frequency range between $a/\lambda=0.652$ and $a/\lambda=0.668$ (see the inset in Fig. 3(c), which represents all-angle SC property), the values of GVD and TOD are obtained as $42.29 (a/2\pi c^2) - 390 (a/2\pi c^2)$ and $5.2 \times 10^3 (a^2/4\pi^2 c^3) - 9 \times 10^4 (a^2/4\pi^2 c^3)$, respectively.
As seen, GVD gets a relatively small value, which enables the generation of the collimated beam. Although an incident beam is collimated during propagating the PC structure, the shape of the beam is distorted due to the relatively larger value of TOD.
As a result, it can be reported that introducing low symmetry reveals an all-angle SC effect in the hexagonal PC structure.
As known, the TOD takes small values when the variation of GVD with the frequency is smooth and nearly linear. Therefore, by adjusting the index contrast or adding more auxiliary PC rods, the GVD as well as the TOD values can be set appropriately. For this reason, we increase the number of auxiliary rods from single to double in the proposed structure, thus the PC transforms into a $C_2$ symmetry structure.

The introduced double auxiliary rods are placed at 0° and 180° with respect to the main PC rod in the counterclockwise (CCW) direction in the unit cell, as seen in Fig. 1(g). The distance between the main rod and auxiliary rods is taken as$d=0.35a$. The radii of the auxiliary rods are fixed to $r=0.1a$. The EFCs of the \nth{3}, \nth{4}, and \nth{5} TM polarization bands are also calculated and presented in Figs. 4(a)-(c), respectively. The schematics of the corresponding PC unit cell are given as insets in the corresponding figures. The frequency contours calculated for the \nth{3} TM band exhibit curved nonlinear shapes. On the other hand, the linear contours are obtained provided by the EFC of the \nth{4} and \nth{5} bands, as shown in Figs. 4(b) and 4(c), respectively.
As expected, a
change in the overall index contrast of the PC unit cell with introduce double auxiliary rods plays an important role in revealing the all-angle SC effect.. These auxiliary rods located at 0° and 180° angles with respect to the main PC rod serve as a physical connection between the main rods in the PC structure which enables the incident light to propagate straightly. The perfect linear contour of the incident light at all angles is observed at the frequency range between  $a/\lambda=0.616$ and $a/\lambda=0.656$ as shown in figure 4(b). This indicates that the PC structure with $C_2$ group symmetry yields the all-angle SC effect at the frequency interval with the broad bandwidth of $\Delta\omega/\omega_c=6.3\%$, which is $2.63$ times greater than the bandwidth of $C_1$ group symmetry structure. Moreover, the EFC of the \nth{5} TM band given in Fig. 4(c) presents an all-angle SC effect for the frequency range between $a/\lambda=0.712$ and $a/\lambda=0.760$ with a bandwidth of $\Delta\omega/\omega_c=6.5\%$. Therefore, we can say that the $C_2$ symmetry structure operates as an all-angle self-collimator at two different frequency bands with wider frequency bandwidths compared to $C_\infty$ and $C_1$ symmetry structures.

To increase the index contrast even more, the number of auxiliary rods is increased to three and placed at the angles of 30°, 150°, and 270° with respect to the main PC rod in a counterclockwise (CCW) direction constituting the $C_3$ symmetry group structure, as shown in Fig. 1(h). Here, opto-geometrical parameters used for the $C_\infty$, $C_1$, and $C_2$ structures are also valid for the $C_3$ group. The EFCs calculated for the \nth{3}, \nth{4}, and \nth{5} TM polarization bands of the $C_3$ symmetry group structure are given in Figs. 4(d)-(f), respectively.However, this structure doesn’t yield any sign of a SC effect as can be inferred from the EFCs because the $C_3$ group symmetry is more complex and sensitive to scattering and diffraction compared to the $C_2$ and $C_\infty$, $C_1$ symmetry groups. Additionally, the EFC behavior is also affected by the degree of freedom in the unit cell of the PC.  Due to the lower degrees of freedom in $C_3$ symmetry in the hexagonal lattice than in $C_1$ and $C_2$ symmetry, light manipulation becomes more difficult, as discussed by the previous studies \cite{erim19,Gumus18_2,gumus_2018_2}. As a result, the $C_3$ group exhibits a complex/nonlinear shaped frequency response that is reflected in the EFCs.

To qualitatively evaluate the all-angle SC characteristics of the designed $C_2$ symmetry group, corresponding steady-state intensity distributions are calculated and presented in Figs. 5(a)-5(c) and Figs. 5(d)-5(f) for the \nth{4} and \nth{5} TM polarization bands, respectively. The results of the steady-state intensity field profiles calculated for the operating frequencies of $a/\lambda= 0.62$, $a/\lambda= 0.63$, and $a/\lambda= 0.64$ are given in Figs. 5(a)-(c), respectively.  It is important to note that these frequencies locate within the all-angle SC frequency range of $a/\lambda= 0.616-0.656$ (see Fig. 4(b), EFCs for the \nth{4} TM band). From the intensity distributions, one can observe light propagation with negligible broadening while keeping a strong confinement of the energy at the center of the structure. Even though the light propagates without spatial expansion, the light is not propagating inside the PC structure by keeping intact the spatial width of the source. In other words, light propagates inside the structure by condensing a large amount of energy at the center of the PC accompanied by the trace of the residual energy wave. On the other hand, in the case of the \nth{5} band, the steady-state intensity profiles of the propagating wave operating at frequencies of $a/\lambda= 0.73$, $a/\lambda= 0.74$, and $a/\lambda= 0.75$ exhibit a less scattered distribution of the collimated beam, as shown in Figs. 5(d)-(f), respectively.

Similarly, the values of GVD and TOD are also calculated for the \nth{4} and \nth{5} bands as given in Figs. 5(g) and 5(h), respectively. 
In these figures, it can be observed that the GVD and TOD values vary between $7.3 \left(\frac{a}{2\pi c^2}\right) - 254.3 \left(\frac{a}{2\pi c^2}\right)$ and $449.2 \left(\frac{a^2}{4\pi^2c^3}\right) - 1.3 \times 10^5 \left(\frac{a^2}{4\pi^2c^3}\right)$ for the TM \nth{4} band, respectively, and between $182.5 \left(\frac{a}{2\pi c^2}\right) - 71.3 \left(\frac{a}{2\pi c^2}\right)$ and $-24380 \left(\frac{a^2}{4\pi^2c^3}\right) - -9619 \left(\frac{a^2}{4\pi^2c^3}\right)$ for the TM \nth{5} band values.
Here, GVD and TOD values are smaller compared to those values in $C_1$ symmetry group case. On the other hand, the variation of GVD values in the \nth{4} band is much more noticeable compared to the variation of GVD in the \nth{5} band. It shows that $C_2$ symmetry group PC structure is more sensitive to the frequency change at the \nth{4} band, which can be observed from steady-state electric field intensity profiles given in Figs. 5(a)-5(c).

It is important to note that the designed PC structure should present efficient optical performance as well as high compatibility with state-of-the-art fabrication techniques. 
Here, $C_2$ symmetry group structure presents the desired all-angle SC characteristics working in a broadband frequency range. However, the $C_2$ group symmetry structure with the isolated auxiliary rods and the main rods can be challenging for high-precision fabrications \cite{Cheng1997}.
On the other hand, if we look at the $C_2$ group symmetry, one can observe that auxiliary rods around the main rod ($\theta=0^\circ$) can be placed in a very close distance (auxiliary rods can touch the main rod) to the main rod. This situation can be beneficial in terms of designing a composite/hybrid structure that can exhibit similar optical characteristics.
The composite PC structure generated using $C_2$ group symmetry ($\theta=0^\circ$) is shown in Fig. 6(a), where two auxiliary rods are replaced by rectangular photonic wire with a width equal to the diameter of auxiliary rods and the same refractive index.
Here, the photonic wire acts as a connecting bridge between isolated main rods and does not violate the $C_2$ group symmetry.
As a result, the hybrid structure resembles $C_2$ group symmetry and is considerably more robust against possible fabrication precision issues. There may be deformation in the radius of rods in the hybrid PC structure fabricated. However, the EFC behavior of the hybrid PC structure is less sensitive to fabrication defects.
As depicted in Figs. 6(b) and 6(c), this hybrid structure exhibits all-angle SC characteristics at the \nth{4} and \nth{5} TM bands,  as shown in  $C_2$ group symmetry PC structure.
The hybrid PC structure exhibits near-zero GVD and TOD performance as shown in Fig. 6(d), which can be used to tune the electromagnetic pulse structure to maintain signal fidelity. 
Furthermore, these near-zero values can be used for further data compression or energy transfer of the wave packet. In addition, the combination of negative TOD with near-zero values can be utilized to shape the wave packet. The hybrid structure also inherits the significant SC potential of the two auxiliary rod structures, as seen in both the EFC in figures 6(b) and (c).

Corresponding steady-state electric field intensity distributions for the \nth{4} and \nth{5}5thTMpolarization bands
are calculated to qualitatively evaluate the all-angle SC properties of the designed hybrid symmetry structure and presented in figure 7. The steady-state intensity field profiles for the selected frequencies operating at $a/\lambda= 0.665$, $a/\lambda= 0.675$, $a/\lambda= 0.685$, $a/\lambda= 0.695$, $a/\lambda= 0.705$, and $a/\lambda= 0.715$ are demonstrated in Figs. 7(a)-(f), respectively. Figure 7(a) provides a detailed representation of the electric field density distributions within the inset structure. Here the image is enlarged by $250\%$.  It is important to note that these frequencies are within the all-angle SC frequencies of $a/\lambda= 0.648-0.736$ as noticed in Fig. 6(c). As a result, the proposed hybrid PC structure demonstrates an all-angle SC effect with negligible broadening and spatial dispersion while ensuring strong confinement of energy at the center line of the structure.

It is also interesting to compare the dispersion performances of the structures proposed in this study with previous theoretical results. Table 1 presents the numerical comparison of GVD, TOD, and bandwidth characteristics at a defined operating frequency range of present work with the studies in literature. As can be observed from the table 1, introducing an auxiliary rod to the highly symmetric hexagonal structure decreases the GVD and TOD values while increasing the bandwidth value. In fact, when we transformed the low-symmetry structure into a hybrid structure, these properties were significantly enhanced, as shown in table 1. Moreover, the SC properties of all our proposed PC structures, except for bandwidth, are comparable with those of structures modified by the square lattice with one auxiliary rod \cite{gumus2018,gumus_2018_2}, two auxiliary rods \cite{Gumus18_2}, and a hybrid structure \cite{Chung:11,Cicek2011}, as well as those without an auxiliary rod \cite{GIDEN2013,Gumus_2020,Zhou2008,Cicek2011}. As a result, whether the hybrid structure is designed from a hexagonal or square lattice, a wider frequency range, lower GVD and TOD values, and wider bandgap properties are achieved, indicating that the collimation properties are significantly enhanced.

\section{Conclusion}

In this study, we introduce a PCstructure with a hexagonal lattice, where adjustments in unit cell symmetry reveal the all-angle SC effect. By optimizing opto-geometric parameters, such as the rotational angle of auxiliary rods and adjacent distances, we conduct a thorough analysis of the SC property, utilizing GVD and TOD characteristics. Furthermore, we explore the interplay between symmetry properties and their impact on dispersion characteristics. Symmetry manipulation affords us a profound understanding of the underlying mechanisms governing light collimation and confinement in these configurations. The PC structure with a $C_1$ symmetry group exhibits an all-angle SC effect within a normalized frequency range of $a/\lambda=0.652$ to $a/\lambda=0.668$, featuring a bandwidth of $\Delta\omega/\omega = 2.4\%$. Breaking symmetry further by transitioning from $C_1$ to $C_2$ group symmetry enhances the SC bandwidth to $\Delta\omega/\omega=6.5\%$. It unveils perfect linear EFCs at two frequency bands: all-angle SC between $a/\lambda=0.616$ and $a/\lambda=0.656$ in the \nth{4} TM band and between $a/\lambda=0.712$ and $a/\lambda=0.760$ in the \nth{5} TM band. In this context, we present and analyze GVD and TOD values within specific ranges for both the TM \nth{4} and \nth{5} bands to develop our understanding of these dispersion characteristics. Additionally, we propose a hybrid PC structure resembling $C_2$ group symmetry, where rectangular photonic wires, matching the diameter of auxiliary rods and sharing the same refractive index replace two auxiliary rods. This hybrid structure features an all-angle SC effect with an operating bandwidth of $\Delta\omega/\omega=11.7\%$, which has near-zero GVD and TOD values. Reducing the symmetry of the PC structure results in lower GVD and TOD values, which contribute to dispersion engineering by improving SC properties. To the best of our knowledge, this study provides the first comprehensive analysis of all-angle SC characteristics in low-symmetry PC structures arranged in hexagonal lattices by using the EFC engineering approach.

The optimized structure proposed in this study may guide the studies aiming innovative design for different applications and can be integrated into new-generation applications requiring light manipulation, such as signal processing in telecommunications, solar cells, and display technologies. The feasibility of the proposed PC structure may be verified by future experimental studies.

\section{Acknowledgement}

This work was supported by Pamukkale University - Scientific Research Project Unit under Project No: BAP-2021FEBE040 and the Scientific and Technological Research Council of Turkey (TUBITAK) under Project No. 118E954. ZMY0 and HO acknowledges YÖK(Council of Higher Education) 100/2000 Project Program for financial support. The authors would like to thank Assoc. Prof. Dr Sami Sozuer for valuable discussions on generating of EFC for hexagonal PC structures.

\FloatBarrier
 \begin{figure}[H]
    \centering
        \includegraphics[width=\textwidth]{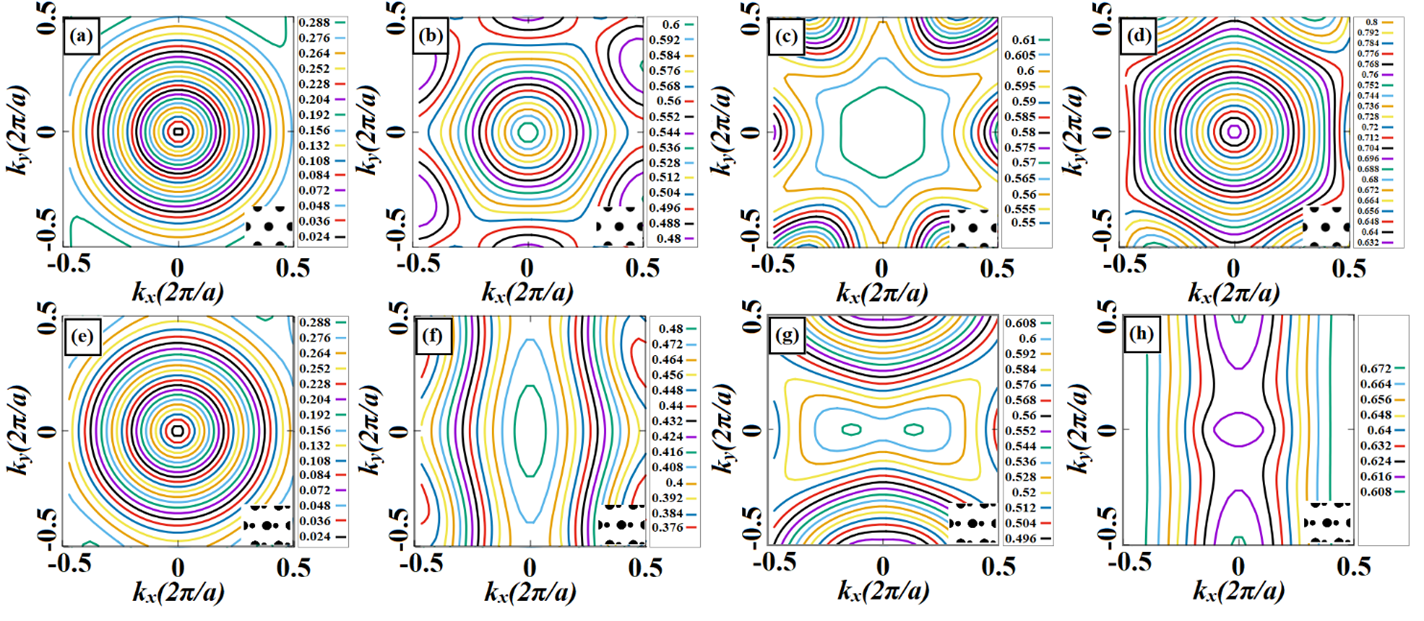}    
    \caption{EFC plots of the hexagonal PC structure without auxiliary dielectric rod ($C_{\infty}$ symmetric structure) for (a) \nth{1}, (b) \nth{2}, (c) \nth{3}, and (d) \nth{4} TM bands. EFCs for hexagonal PC structure with a single auxiliary dielectric rod ($C_1$ symmetric structure) for (a) \nth{1}, (b) \nth{2}, (c) \nth{3}, and (d) \nth{4} TM bands.}
  \label{fig:fig2}
\end{figure}
\FloatBarrier
 \begin{figure}[H]
    \centering
        \includegraphics[width=\textwidth]{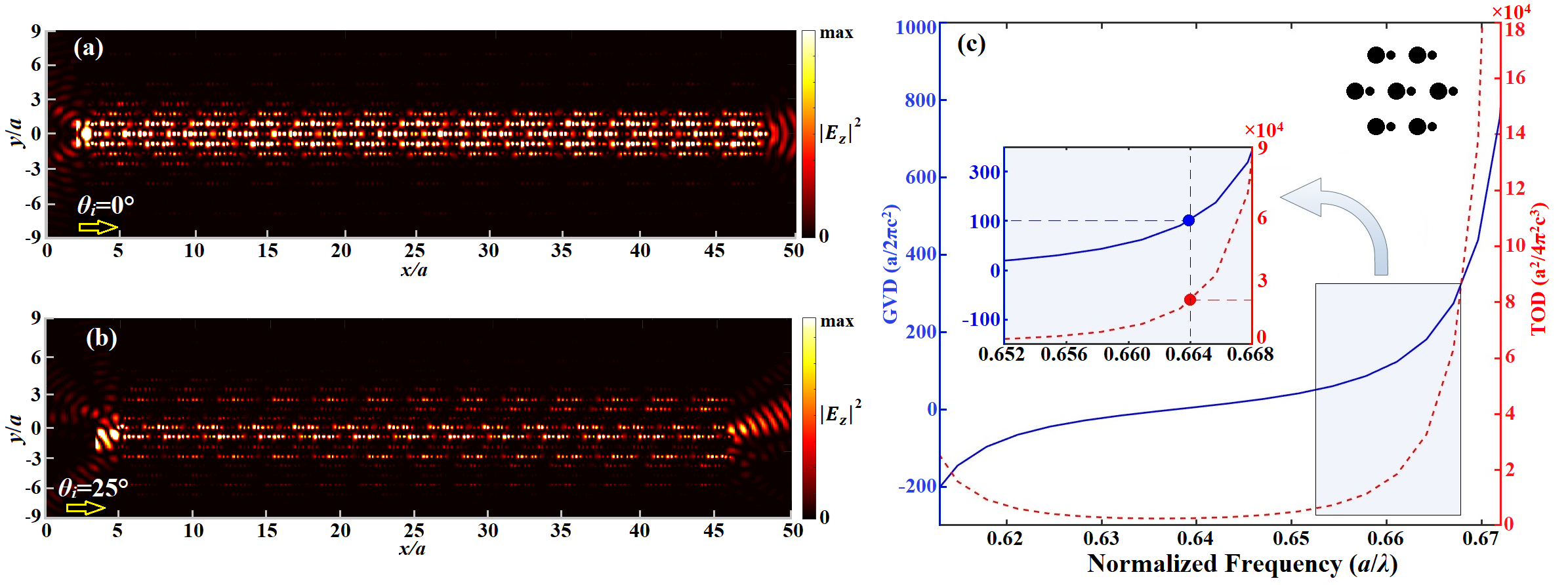}
 \caption{Steady-state electric field intensity distributions of the $C_1$ symmetry group PC structure operating at a frequency of $a/\lambda=0.65$ under (a) 0° and (b) 25° incident light illumination. (c) Variation of the GVD and TOD values for the $C_1$ symmetry group PC structure as a function of normalized frequency, with a representation of the unit cell schematic as an inset.}
  \label{fig:fig3}
\end{figure}
\FloatBarrier
 \begin{figure}[H]
    \centering
        \includegraphics[width=\textwidth]{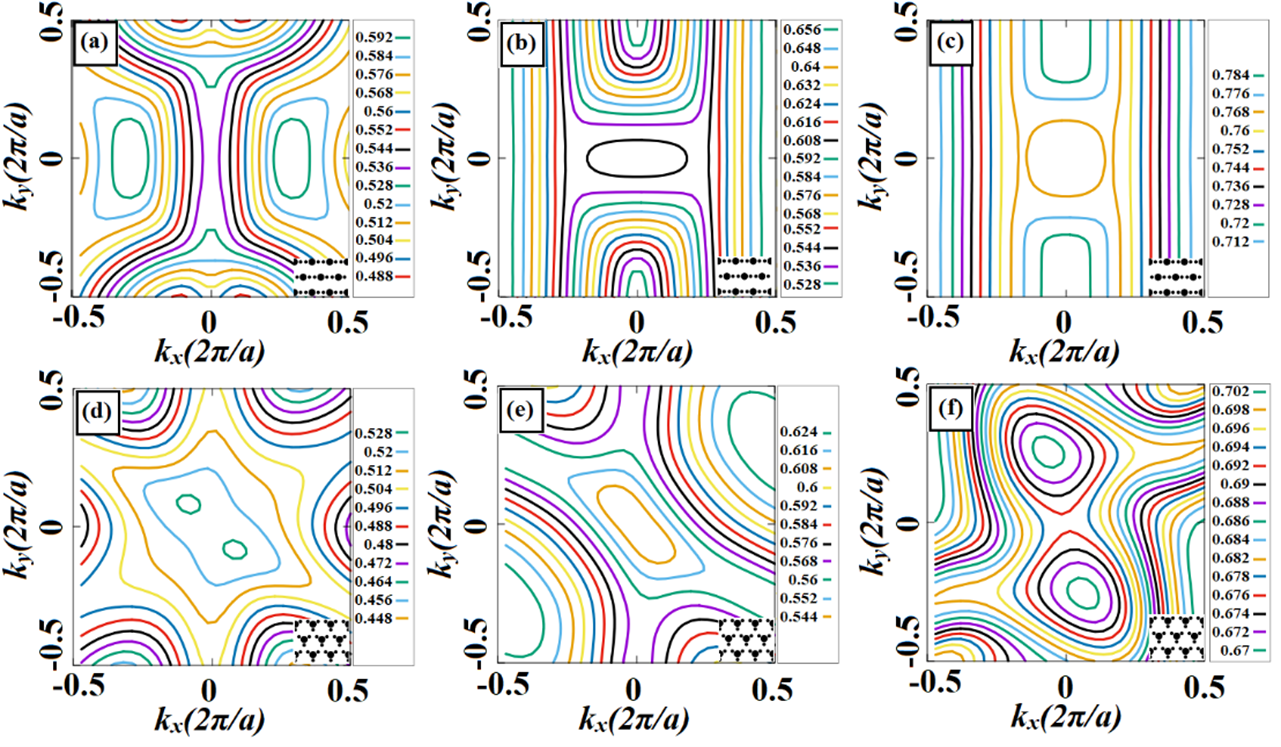}
    \caption{EFC plots of the hexagonal PC structure with two dielectric rods ($C_2$ symmetric structure) for (a) the \nth{3}, (b) the \nth{4} and (c)the \nth{5} TM bands. EFCs for the hexagonal PC structure with three auxiliary dielectric rods ($C_3$ symmetric structure) for (a) the \nth{3}, (b) the \nth{4} and (c) the \nth{5} TM bands. }
  \label{fig:fig4}
\end{figure}
\FloatBarrier
 \begin{figure}[H]
    \centering
        \includegraphics[width=\textwidth]{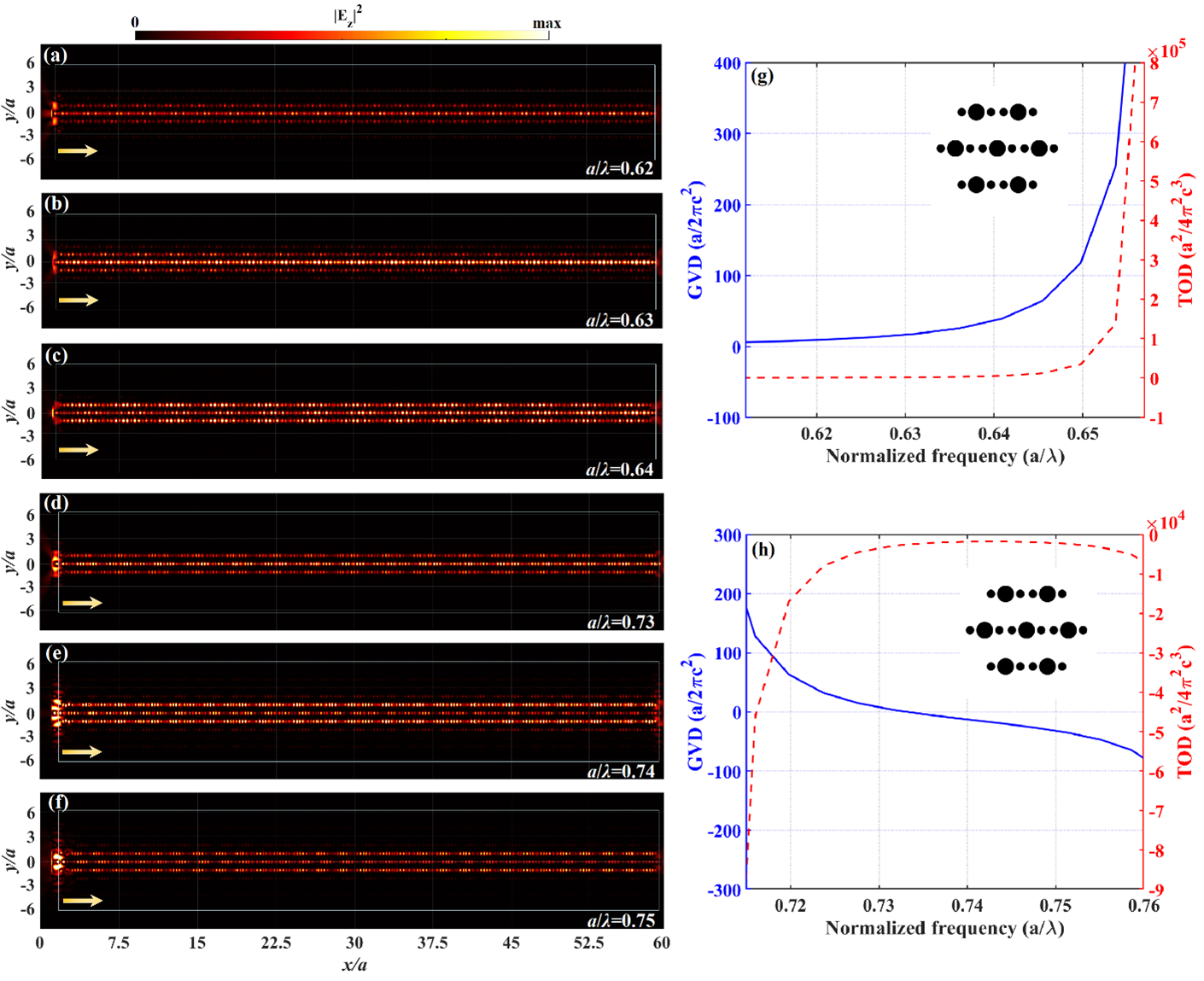}
   \caption{Steady-state electric field intensity distributions of $C_2$ group symmetry PC structure operating at the frequencies of (a) $a/\lambda=0.620$, (b) $a/\lambda=0.630$, and (c) $a/\lambda=0.640$ in the \nth{4} TM band. Steady-state electric field intensity distributions of the same structure operating at \nth{5} TM band of (d) $a/\lambda=0.730$, (e) $a/\lambda=0.740$, and (f) $a/\lambda=0.750$. Variation of the GVD and TOD values for the $C_2$ symmetry group PC structure as a function of normalized frequency at (g) the \nth{4} and (h) the \nth{5} TM bands, with representation of unit cell schematic as an inset.}
  \label{fig:fig5}
\end{figure}
\FloatBarrier
 \begin{figure}[H]
    \centering
        \includegraphics[width=0.8\textwidth]{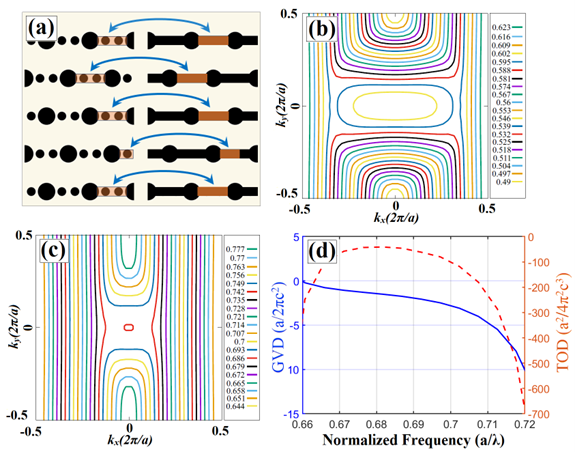} 
    \caption{(a) Schematical view of the proposed the hybrid PC structure transformed/adapted from $C_2$ group symmetry structure. EFC diagrams of the hybrid structure for (b) the \nth{4} and (c) the \nth{5} TM bands. (d) Variations of the GVD and TOD values for the hybrid PC structure as a function of normalized frequencies at the \nth{5} TM bands.}
  \label{fig:fig6}
\end{figure}
\FloatBarrier
 \begin{figure}[H]
   
        \includegraphics[width=\textwidth]{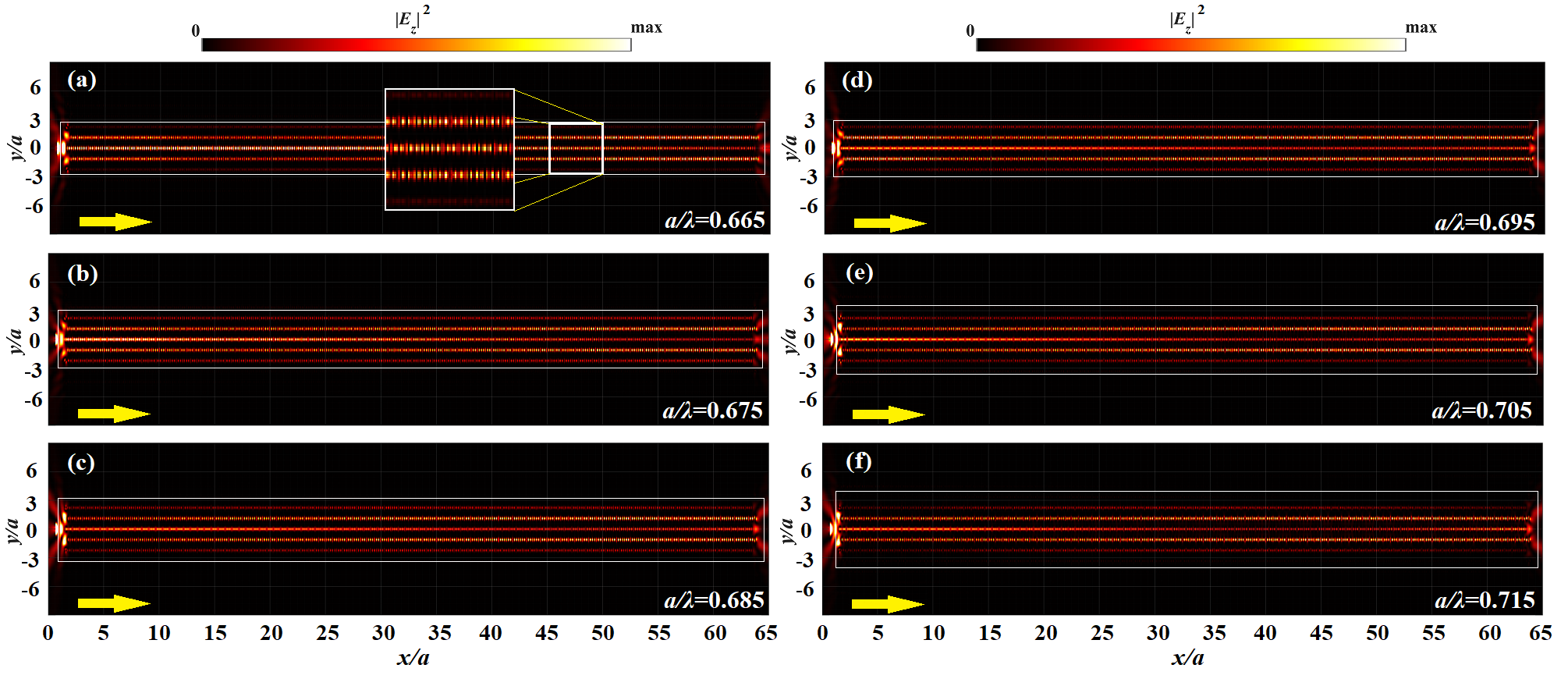}
   \caption{Steady-state electric field intensity distributions of hybrid PC structure operating at the frequencies of (a) $a/\lambda=0.665$, (b) $a/\lambda=0.675$, (c) $a/\lambda=0.685$, (d) $a/\lambda=0.695$, (e) $a/\lambda=0.705$, and (f) $a/\lambda=0.715$ in the \nth{5} TM band.The inset in tab (a) is a close-up view of the electric field intensity distributions within the structure. }

  \label{fig:fig7}
\end{figure}
\FloatBarrier

\begin{table}[ht]
\centering
\begin{adjustbox}{max width=\textwidth}
\begin{tabular}{lllll}
\hline
\textbf{Study} &
  \textbf{Type of the PC Structure} &
  \textbf{Frequency range} $\left( \frac{a}{\lambda} \right)$ &
  \textbf{GVD} $\left( \frac{a^2}{2\pi c^2} \right)$ &
  \textbf{Bandwidth} $\left( \frac{\Delta \omega}{\omega_c} \right)$ \\ \hline
\multirow{3}{*}{Proposed study} &
  Hexagonal lattice, low symmetry (one aux. rod) &
  $0.652--0.668$ &
 $ 42.3--390$ &
  $2.4\% $\\
     & Hexagonal lattice, low symmetry (two aux. rods)  & $0.712--0.760$ & $181.5--71.3$   & $6.5\%$  \\
     & Hexagonal lattice, low symmetry hybrid structure & $0.648--0.736$ &$ 0--10$         & $11.7\%$ \\
\cite{gumus2018} & Square lattice, low symmetry (one aux. rod)      & $0.610--0.680 $& $-100$--$100$ &$ 11\%$   \\
\cite{gumus_2018_2} & Square lattice, low symmetry (one aux. rod)      &$ 0.610--0.635$ & $-59--0$      &$ 4.1\%$  \\
\cite{Gumus18_2} & Square lattice, low symmetry (two aux. rods)     &$ 0.570--0.660$ & $-0.6--24.9$  & $15\%$   \\
\cite{GIDEN2013} & Square lattice, high symmetry (star rods)        & $0.5405$       & $-0.0904$     & $16.4\% $\\
\cite{Gumus_2020} & Square lattice, high symmetry                    & $0.481--0.701$ & $0.02--200$     & $37\%$   \\
\cite{Cicek2011} & Square lattice, high symmetry                    & $0.59$         & $\sim 0.003$  & $39\%$   \\
\cite{Zhou2008} & Square lattice, high symmetry                    & $0.2915$       & $0 $            & $3.6\%$  \\
\cite{Chung:11} & Square lattice, high symmetry, hybrid structure  & $0.46--0.52$   & $-2--1.2$     & $4.5\% $ \\ 
\hline
\caption{GVD, TOD and bandwidth values of PC structures proposed in this study, along with those of the previous computational studies.}
\end{tabular}
\end{adjustbox}
\end{table}
\FloatBarrier
\newpage
\bibliographystyle{unsrtnat}

\bibliography{bibi.bib}

\end{document}